\documentstyle[12pt]{article}
\if@twoside  m
\else
\fi
   \newcommand{\ie}{{\em i.e.}}
   \newcommand{\eg}{{\em e.g.}}
   \newcommand{\rhs}{{\em rhs }}
   \newcommand{\R}{I\!\!R}
   \newcommand{\OO}{{\cal O}}

\begin{document}
\title{}
\date{}
\author{}
\noindent
{\Large\bf Point interaction in dimension two and three as models of
small scatterers}
\vspace{15mm}
\begin{quote}
{\large P.~Exner and P.~\v{S}eba}
\vspace{3mm}

{\em Nuclear Physics Institute, Academy of Sciences \\
CZ--25068 \v Re\v z near Prague, and \\ Doppler 
Institute, Czech Technical University, \\ B\v rehov\'a 7, 
CZ--11519 Prague, Czech Republic}
\vspace{10mm}

In addition to the conventional renormalized--coupling--constant
picture, point interactions in dimension two and three are shown to
model within a suitable energy range scattering on localized
potentials, both attractive and repulsive. 
\end{quote}
\vspace{10mm}

Point interactions described by formal $\,\delta$--shaped potentials
were used in quantum mechanics from the early thirties. It lasted
several decades, however, until a proper way to handle the
corresponding Schr\"odinger operators was found. Following the
observation of Berezin and Faddeev \cite{BF}, the problem was studied
systematically in the eighties. The results are summarized in the
monograph \cite{AGHH}, references to some recent work are given, \eg,
in \cite{AESS,EGST}. In its present form the point--interaction method
represents a versatile tool for constructing solvable models whose
power has not been yet, to our opinion, appreciated fully in the
physical community.

Within the standard quantum mechanical formalism point interactions
can be constructed as long as the configuration space dimension does
not exceed three, and they can be interpreted as limits of suitable
families of squeezed potentials. In the one--dimensional case the
limiting argument admits a straightforward interpretation, because
the point--interaction coupling constant is just the integral of the
approximating potential: a slow particle on the line with a
well--spread wavefunction ``sees" only the average value of a
localized potential.

In distinction to that the approximation by scaled potentials in
dimension two and three requires existence of a zero--energy
resonance and a particular coupling--constant renormalization; a
detailed discussion of the related mathematical problems can be found
in \cite{AGHH,GHM}. This gives these point interactions a flavour of
something exceptional, for instance, one is led to believe that they
cannot be used to model well--localized {\em repulsive} potentials.
That would have consequences, in particular, for recent studies of
quantum wires with natural or artificial impurities
\cite{CBC1,CBC2,EGST,KSF,MNB}. 

To be more specific, recall that the Hamiltonian with $\,N\,$ point
interactions is formally given by 
   \begin{equation} \label{H}
H\,=\,-\,\Delta \,+\, \sum_{i=1}^N\, \alpha_i 
\delta(\vec x \!-\!\vec x_i)\,, 
\end{equation}
where $\,\Delta\,$ denotes the Laplace operator of appropriate
dimension and $\,\alpha_i,\;\vec x_i\,$ are the coupling constants
and the localization of the point interactions, respectively. If a
particular $\,\alpha_i\,$ in the above expression is positive, the
corresponding point interaction is interpreted as ``repulsive" and
vice versa --- see, \eg, \cite{CBC1,CBC2}. 

Our aim here is to show, that such an understanding of (\ref{H}) is
misleading, at least if we want to identify the formal coupling
constants $\,\alpha_i\,$ with the parameters in the boundary
conditions which define the point interactions rigorously
\cite{AGHH}. In fact ---- as we are going to demonstrate by
approximating the point interactions by a suitable well localized
potential --- only the negative coupling constants $\,\alpha_i\,$ can
be {\em typically} given a straightforward physical interpretation,
and moreover, it is not appropriate to associate $\,\alpha_i<0\,$
with an attractive interaction, since the corresponding scattering
matrix describes at low energies both attractive as well as repulsive
potentials.

Let us stress that we speak about approximation by a fixed potential.
If we start squeezing it we arrive at the free Hamiltonian unless a
special choice of the potential and the coupling--constant
renormalization is made \cite{AGHH}. Nevertheless, we shall see that
a point interaction represents a natural model for the scattering
behaviour of a well localized potential within a certain energy
interval. 

In what follows we focus on one point interaction only. Due to the
zero radius of the potential the only nontrivial contribution
to the on--shell S--matrix is given by its s--wave component
$\,S_0=e^{2i\delta_0}\,$ with the phase shift $\,\delta_0\,$ given by
the formulae \cite[Secs. I.1, I.5]{AGHH} 
   \begin{equation} \label{shifts}
\delta_0(k)\,=\,\left\{
\begin{array}{ll}
\arctan\left(\frac{\pi}{2(2\pi\alpha+\gamma+ln\frac{k}{2})}\right) 
& \mbox{\qquad in dimension 2}\\
& \\
\arctan\left(\frac{k}{4\pi\alpha}\right) & \mbox{\qquad in dimension 3}
\end{array} \right.
   \end{equation} 
where $\,k^2\,$ is the energy of the scattered particle and
$\gamma=0.577..$ is the Euler's constant.

Let us start with the two--dimensional case. A particular case of the
problem we address here has been recently discussed in connection with 
semiclassical resonances in scattering on a system of hard planar
discs \cite{RWW}. The authors of this paper derive the s--wave
scattering amplitude for a disc of radius $\,a\,$ to be
   \begin{equation} \label{RWW amplitude}
d\,\,\approx\, {2\pi\over \ln\left(2\over ka\right)-\gamma+\, {\pi
i\over 2}}
   \end{equation}
for $\,ka\ll 1\,$. A comparison with (\ref{shifts}) shows that we
will obtain the same phase shift if we put the corresponding ``coupling
constant" equal to
   \begin{equation} \label{identification2}
\alpha\,=\, {1\over 2\pi}\, \ln a\;.
   \end{equation}
To illustrate that this is true not only for hard discs but also for
well--distinguished potential barriers and {\em nonresonant} wells,
we shall discuss below the low--energy scattering on a
compactly supported potential; to avoid technicalities we shall
consider just the example of a rectangular barrier/well.

Consider therefore a potential in $\,\R^2\,$ which equals $\,V_0\,$
within a circular disc of diameter $\,a\,$ and zero otherwise. Due to
the rotational symmetry, the scattering may be considered in each
partial wave separately. The corresponding solutions to the radial
Schr\"odinger equation of energy $\,k^2\,$ are well known,
   \begin{equation} \label{radial solution2} 
f(r)\,=\, \left\lbrace\;
\begin{array}{lcc} c\,\sqrt{r}\, J_{\ell}(\kappa r) & \quad \dots
\quad & |r|<a \\ \\
\sqrt{r}\, \left\lbrack\, J_{\ell}(kr)+ ie^{i\delta_{\ell}} \sin
\delta_{\ell}\, H_{\ell}^{(1)}(kr) \right\rbrack & \quad \dots \quad
& |r|>a  \end{array} \right.
   \end{equation}
where $\,\kappa:=\sqrt{k^2\!-V_0}\,$.

In the hard--disc case, $\,V_0=\infty\,$, the inner solution is
absent, $\,c=0\,$, and the condition $\,f(a)=0\,$ yields easily
   \begin{equation} \label{disc phase}
\delta_{\ell}(k)\,=\, \arctan\, {J_{\ell}(ka)\over Y_{\ell}(ka)}\;. 
   \end{equation}
At high energies, these phase shifts behave as $\,-ka+\,{\pi\over
2}(2\ell\!-\!1)\,$. On the other hand, for $\,ka\ll 1\,$ we have
$$
\delta_{\ell}(k)\,\approx\, -\,\arctan\, {1\over\pi}\left( 2\over
ka\right)^{2\ell} 
$$
if $\,\ell\ne 0\,$, which tends to the ``free" value $\,{\pi\over
2}\; ({\rm mod\,}\pi)\,$ as $\,a\to 0\,$, while the s--channel
asymptotics is
   \begin{equation} \label{s-channel delta}
\delta_0(k)\,\approx\, \arctan\, {\pi\over 2\left(\gamma+\ln{ka\over
2} \right)}
   \end{equation}
up to $\,\OO\left((ka)^{-2}\right)\,$. Since the point--interaction
is given by expression (\ref{shifts}) with the denominator
$\,2\left(2\pi\alpha+ \gamma+\ln{k\over 2} \right)\,$, we arrive
again at the identification (\ref{identification2}).

For a finite barrier/well, $\,|V_0|<\infty\,$, the requirement of
smooth matching of the inner and outer solution in (\ref{radial
solution2}) leads to 
   \begin{equation} \label{barrier phase2}
\delta_{\ell}(k)\,=\, {\pi\over 2}\,-\, \arg\left\lbrace
\left\lbrack\, k\,J_{\ell}(\kappa a){H_{\ell}^{(1)}}'(ka)- \kappa\,
J'_{\ell}(\kappa a)H_{\ell}^{(1)}(ka) \right\rbrack \left(
\kappa\over|\kappa|\right) ^{\ell} \right\rbrace\,.
   \end{equation}
We are interested in the behaviour of this function for
 $\,|\kappa|a
\gg 1\gg ka\,$. The square bracket is then mostly dominated by the
second term. An exception may occur in the well case ($\,V_0<0\,$)
when $\,\kappa\,$ is real and the first--kind Bessel function has an
oscillatory asymptotics. The second--term dominance is then violated
in the vicinity of the points where $\,\kappa a\approx\, {\pi\over
4}\, (4n\!+\!2\ell\!+\!1)\,$, \ie, resonances of the corresponding
scattering process.

Away of the resonances, the dominating role of the second term is
checked by estimating the ratio
$$
{k\,{H_{\ell}^{(1)}}'(ka)\over \kappa\, H_{\ell}^{(1)}(ka)}\,=\,
{\ell\over\kappa a}\,-\, {k\,H_{\ell+1}^{(1)}(ka)\over \kappa\,
H_{\ell}^{(1)}(ka)}\,.
$$
If $\,\ell\ne 0\,$ and $\,ka\ll 1\,$, the \rhs behaves as
$\,\ell/2\kappa a\,$, while for $\,\ell=0\,$ we have instead
$\,-(\kappa a\,\ln ka)^{-1}\,$; both are $\;\ll 1\,$ under our
assumption. Furthermore, $\,\kappa J'_{\ell}(\kappa a)\,$ is real if 
$\,\kappa\in\R\,$ or if $\,\kappa\,$ is purely imaginary and
$\,\ell\,$ is even. If $\,\ell\,$ is odd, $\,\kappa J'_{\ell}(\kappa
a)\,$ is purely imaginary; since the same is true for
$\,(\kappa/|\kappa|)^{\ell}$, we find in all cases
$\,\delta_{\ell}(k) \approx\,-\,{\pi\over 2}\,+\arg
H_{\ell}^{(1)}(ka)\,$, or
   \begin{equation} \label{barrier asymptotics}
\delta_{\ell}(k)\,\approx\, \arctan\, {J_{\ell}(ka)\over
Y_{\ell}(ka)} \qquad ({\rm mod\;}\pi)\,.
   \end{equation}

The three--dimensional situation can be treated in a similar way. 
The radial solutions in the $\,\ell$--th partial wave are now given
by 
   \begin{equation} \label{radial solution3}
f(r)\,=\, \left\lbrace\; \begin{array}{lcc} c\,r\,
j_{\ell}(\kappa r) & \quad \dots \quad & |r|<a \\ \\
r \left\lbrack\, j_{\ell}(kr)+ ie^{i\delta_{\ell}} \sin
\delta_{\ell}\, h_{\ell}^{(1)}(kr) \right\rbrack & \quad \dots \quad
& |r|>a  \end{array} \right.
   \end{equation}
with the same $\,\kappa\,$ as above. In the hard--ball case the
condition $\,f(a)=0\,$ yields 
   \begin{equation} \label{ball phase}
\delta_{\ell}(k)\,=\, \arctan\, {j_{\ell}(ka)\over y_{\ell}(ka)}\;. 
   \end{equation}
At high energies we have $\,\delta_{\ell}(k)\approx\,-ka+
\,{\ell\pi\over 2}\,$, while the low--energy asymptotics is
$$
\delta_{\ell}(k)\,\approx\, -\, {(ka)^{2\ell+1}\over (2\ell+1)
((2\ell-1)!!)^2} \,+\, \OO(k^{2\ell+3})\,.
$$
In particular, the s--wave expression simplifies to the form
   \begin{equation} \label{s-channel delta3}
\delta_0(k)\,=\, -ka\,.
   \end{equation}
This has to be compared with the point--interaction phase--shift
(\ref{shifts}); they have the same 
low--energy behaviour if we put
   \begin{equation} \label{identification3}
\alpha\,=\, -\, {1\over 4\pi a}\;.
   \end{equation}

For a finite barrier/well (\ref{barrier phase2}) is replaced by
   \begin{equation} \label{barrier phase3}
\delta_{\ell}(k)\,=\, {\pi\over 2}\,-\, \arg\left\lbrace
\left\lbrack\, k\,j_{\ell}(\kappa a){h_{\ell}^{(1)}}'(ka)- \kappa\,
j'_{\ell}(\kappa a)h_{\ell}^{(1)}(ka) \right\rbrack \left(
\kappa\over|\kappa|\right) ^{\ell} \right\rbrace\,.
   \end{equation}
With the exception of the resonance case, $\,\kappa a\approx\,
{\pi\over 2}\,(2n\!+\!\ell\!+\!1)\,$, the second term in the square
bracket dominates for $\,|\kappa|a \gg 1\gg ka\,$, and we have
   \begin{equation} \label{barrier asymptotics3}
\delta_{\ell}(k)\,\approx\, \arctan\, {J_{\ell}(ka)\over
Y_{\ell}(ka)} \qquad ({\rm mod\;}\pi)\,.
   \end{equation}

Concluding the argument, we have found that --- with the exception
of the resonance case --- a square circular barrier/well in dimension
two or three behaves under the condition $\,|\kappa|a \gg 1\gg ka\,$
as a hard disc/ball {\em independently of the sign of} $\,V_0\,$, and
consequently, that its low--energy scattering behaviour can be
modeled by that of the corresponding point interaction with the
coupling constant (\ref{identification2}) or (\ref{identification3}),
respectively. 

It is important that in both cases the value of the coupling constant
depends only on the radius $\,a\,$ of the interaction domain being
independent of the potential strength $\,V_0\,$, provided $\,|V_0|a^2
\gg 1\,$. Moreover, the two--dimensional coupling constant is
negative for $\,a<1\,$. In dimension three, {\em a fortiori,} the
identification (\ref{identification3}) fixes the sign of the coupling
constant for all $\,a\,$; a positive coupling constant can be
achieved only in the resonance case. This means that when point
interactions are used to model various physical situations with
localized scatterers, negative coupling constants are the natural
choice, while positive values have to be reserved for exceptional
resonance states of the system.

In both cases the point interaction with the ``correct" scattering
properties has a bound state \cite{AGHH}, however, the eigenvalue is
well below the energy interval specified by the requirement $\,ka \ll
1\,$. Recall that a similar effect has been observed for the
one--dimensional $\,\delta'$--interaction \cite{AEL} and its
generalization to graphs \cite{E1} whose low--energy behaviour can be
modeled by complicated geometric scatterers; it is more illustrative
than the ``true" approximation with a renormalized coupling constant
and a non--local potential \cite{S2}. The coupling constant
has in this case also a fixed sign and a transparent meaning as a
summary length of the internal links of the scatterer.

\section*{Acknowledgement}
The work has been partially supported by the Grant AS No.148409.


\begin{thebibliography}{article}
   \bibitem{AGHH}
S.~Albeverio, F.~Gesztesy, R.~H\o egh-Krohn, H.~Holden: {\em Solvable Models
in Quantum Mechanics}, Springer, Heidelberg 1988.
   \vspace{-.8em}
   \bibitem{AESS}
J.--P.~Antoine, P.~Exner, P.~\v Seba, J.~Shabani: A mathematical model
of heavy--quarkonia decays, {\em Ann.Phys.} {\bf 233} (1994), 1--16.
   \vspace{-.8em}  
   \bibitem{AEL}
J.E.Avron, P.Exner, Y.Last: Periodic Schr\"odinger operators with
large gaps and Wannier--Stark ladders, {\em Phys.Rev.Lett.} {\bf 72} 
(1994), 896--899.
   \vspace{-.8em}
   \bibitem{BF}
F.A.~Berezin, L.D.~Faddeev: A remark on Schr\"odinger's equation with
a singular potential, {\em Sov.Math.Doklady} {\bf 2} (1961), 372--375.
\vspace{-.8em}
   \bibitem{CBC1}
S.~Chaudhury, S.~Bandyopadhyay, M.~Cahay: Spatial distribution of
current and Fermi carriers around localized elastic scatterers in
quantum transport, {\em Phys.Rev.} {\bf B45} (1992), 11126--11135.
   \vspace{-.8em}
   \bibitem{CBC2}
S.~Chaudhury, S.~Bandyopadhyay, M.~Cahay: Current, potential,
electric field, and Fermi carrier distributions around localized
elastic scatterers in phase--coherent quantum magnetotransport, {\em
Phys.Rev.} {\bf B47} (1993), 12649--12662. 
   \vspace{-.8em}
   \bibitem{E1}
P.Exner: Contact interactions on graph superlattices, {\em J.Phys.}
{\bf A29} (1996), 87--102. 
   \vspace{-.8em}
   \bibitem{EGST}
P.~Exner, R.~Gawlista, P.~\v Seba, M.~Tater: Point interactions in a strip, 
{\em submitted for publication}
   \vspace{-.8em}
   \bibitem{GHM}
A.~Grossmann, R.~H\o egh--Krohn, M.~Mebkhout: A class of explicitly
soluble, local, many--center Hamiltonians for one--particle quantum
mechanics in two and three dimensions, {\em J.Math.Phys.} {\bf 21}
(1980), 2376--2385.
   \vspace{-.8em}
   \bibitem{KSF}
G.~Kirczenow {\em et al.}: Artificial impurities in quantum wires:
from classical to quantum behavior, {\em Phys.Rev.Lett.} {\bf 72}
(1994), 2069--2072.
   \vspace{-.8em}
   \bibitem{KNR}
I.V.~Krive, S.~Naftulin, A.S.~Rozhavsky: Scattering by an
ultralocal potential in a nontrivial topology, {\em Ann.Phys.} {\bf
232} (1994), 225--242.
   \vspace{-.8em}
   \bibitem{MNB}
A.~Mosk, Th.M.~Niewenhuisen, C.~Barnes: Theory of semiballistic wave
propagation, {\em cond--mat/9601075}
   \vspace{-.8em}
   \bibitem{RWW}
P.~Rosenquist, N.D.~Whelan, A.~Wirzba: Small disks and semiclassical
resonances, {\em chao--dyn/9602017}
   \vspace{-.8em}
\bibitem{S2}
P.~\v Seba: Some remarks on the $\,\delta'$--interaction in one
dimension, {\em Rep.Math. Phys.} {\bf 24} (1986), 111--120.
   \vspace{-.8em}

   \end{thebibliography}
   \end{document}